\documentclass[aps,prl,twocolumn,showpacs,superscriptaddress,floatfix,longbibliography,nofootinbib]{revtex4-1}  

\usepackage{graphicx}  
\usepackage{bm}        
\usepackage{amssymb}   
\usepackage{amsmath}   
\usepackage{upgreek}	
\usepackage[utf8]{inputenc} 
\usepackage[english]{babel} %
\usepackage{url}
\usepackage{dcolumn}
\usepackage{bm}
\usepackage{hyperref}
\usepackage{multirow}
\usepackage{stfloats}

\DeclareUnicodeCharacter{00A0}{}
\usepackage[scaled=1.0,helvratio=.94]{newtxtext}    
\usepackage[bigdelims, scaled=1.0,vvarbb,subscriptcorrection]{newtxmath}    
\usepackage[scr=boondoxo,cal=boondoxo]{mathalfa}
\usepackage[activate={true,nocompatibility},final,tracking=true,kerning=true,spacing=true,factor=1000,stretch=20,shrink=20]{microtype}

\makeatletter
\renewcommand*{\fnum@figure}{{\normalfont\bfseries \figurename~\thefigure}}
\renewcommand*{\@caption@fignum@sep}{\textbf{: }}
\makeatother
\usepackage[a4paper,top=2cm,bottom=2cm,left=1.6cm,right=1.6cm]{geometry}

\usepackage{scrextend}
\begin{document} 

\title{Twin-lattice atom interferometry}

\def\affzarm {\affiliation{ZARM, Universit\"at Bremen, Am Fallturm, D-28359 Bremen, Germany }}
\def\affiqo  {\affiliation{Institut f\"ur Quantenoptik, Leibniz Universit\"at Hannover, Welfengarten 1, D-30167 Hannover, Germany }}
\def\affitp  {\affiliation{Institut f\"ur Theoretische Physik, Leibniz Universit\"at Hannover, Appelstr. 2, D-30167 Hannover, Germany }}

\author{M.~Gebbe}
\email[E-mail: ]{gebbe@zarm.uni-bremen.de}
\affzarm
\author{J.-N.~Siem\ss}
\email[E-mail: ]{jan-niclas.siemss@itp.uni-hannover.de}
\affitp
\affiqo
\author{M.~Gersemann}
\affiqo
\author{H.~M\"untinga}
\author{S.~Herrmann}
\author{C.~L\"ammerzahl}
\affzarm
\author{H.~Ahlers}
\author{N.~Gaaloul}
\author{C.~Schubert}
\affiqo
\author{K.~Hammerer}
\affitp
\author{S.~Abend}
\email[E-mail: ]{abend@iqo.uni-hannover.de}
\affiqo
\author{E.~M.~Rasel}
\affiqo

\begin{abstract}
Inertial sensors based on cold atoms have great potential for navigation, geodesy, or fundamental physics. 
Similar to the Sagnac effect, their sensitivity increases with the space-time area enclosed by the interferometer.
Here, we introduce twin-lattice atom interferometry exploiting Bose-Einstein condensates.
Our method provides symmetric momentum transfer and large areas in palm-sized sensor heads with a performance similar to present meter-scale Sagnac devices. 
\end{abstract}

\maketitle

\section*{Introduction}
Atom interferometers measure inertial forces, atomic properties and quantities like the photon recoil or the gravitational constant with high precision and accuracy~\cite{Savoie18SciAdv,Freier16JoP,Asenbaum17PRL,Parker18science,Rosi14Nat}.
Modern fields of application comprise navigation, observation of Earth's gravity and rotation as well as terrestrial and spaceborne gravitational-wave detection in the infrasound domain~\cite{Canuel18SciRep,Schubert19,ELGAR,Hogan11GRG}.
Achieving interferometers with the needed space-time areas remains challenging, though.

We present twin-lattice interferometry as a novel method to form symmetric interferometers featuring matter waves with large relative momentum. 
By employing two counterpropagating optical lattices~\cite{Pagel19arxiv}, we create from a Bose-Einstein condensate (BEC) a Mach-Zehnder type interferometer, made of wave packets with a differential momentum of more than 400 photon recoils, the largest reported so far.
In this way, we realize a Sagnac loop of an area $A=7.6$~mm$^2$ area on a baseline of only 2.43~mm.

\begin{figure}[h]
\begin{center}
\includegraphics[width=\columnwidth]{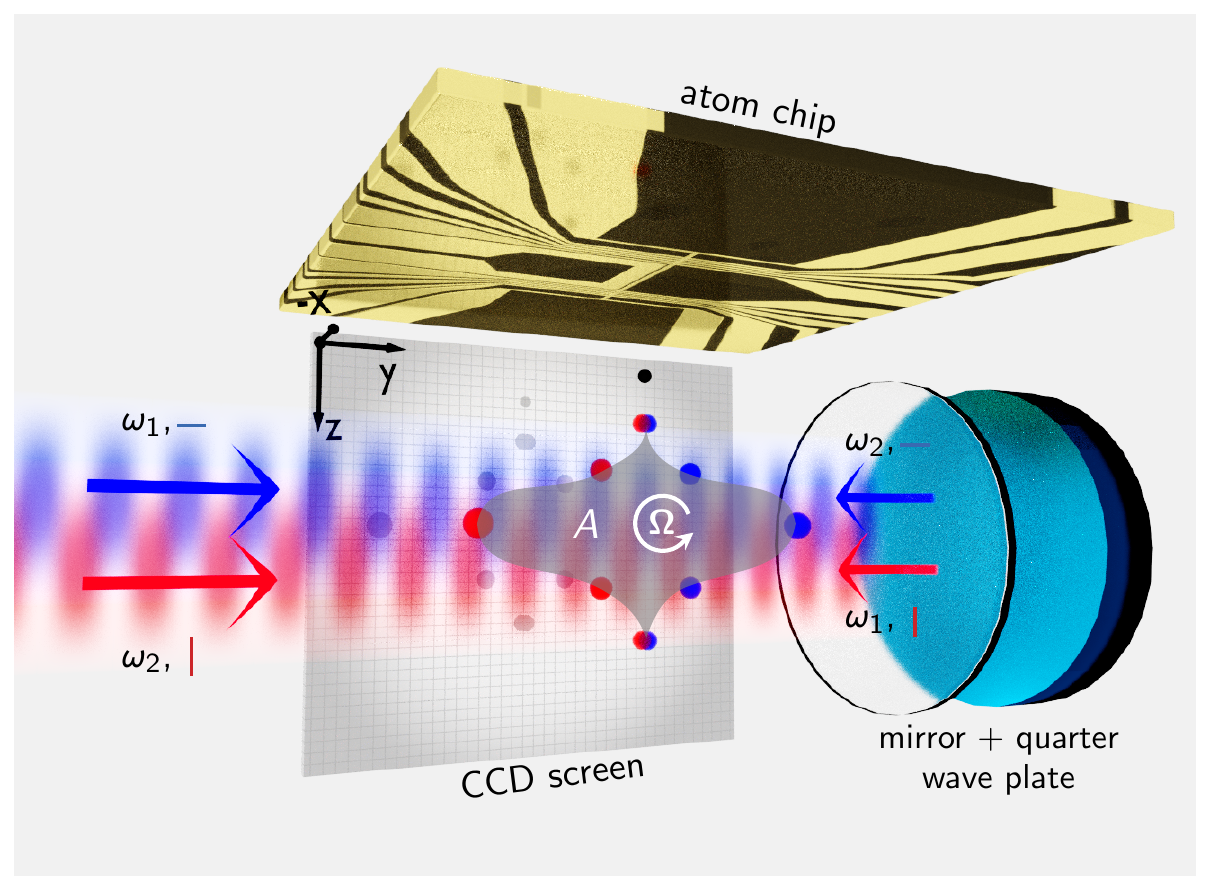}
\caption{ \textbf{Twin-lattice setup.} The twin lattice is formed by retroreflecting light featuring two frequencies with linear orthogonal polarization.
A quarter-wave plate in front of the retroreflector alters the polarization to generate two counterpropagating lattices (indicated in red and blue).
After release from the atom chip and state preparation, the BEC is symmetrically split and recombined by the lattices driving double Bragg diffraction (DBD) and Bloch oscillations (BO).
In this way, the interferometer arms form a Sagnac-loop enclosing an area $A$ (shaded in gray) for detecting rotations $\mathbf{\Omega}$.
The interferometer output ports are detected on a CCD chip by absorption imaging. 
}
\label{fig:setup}
\end{center}
\end{figure}

The sensitivity of atom interferometers towards inertial forces and gravitational waves increases linearly with the differential kinetic momentum and, hence, the latter is exploited as a lever.
Benchmark experiments have so far realized asymmetric and symmetric momentum transfer with Raman diffraction~\cite{Leveque09PRL,Berg15PRL,Jaffe18PRL}, sequential and higher-order Bragg transitions~\cite{Chiow11PRL,Kovachy15Nature,Ahlers16PRL,Plotkin18PRL} as well as Bloch oscillations~\cite{Mueller09PRL,Clade09PRL,Mcdonald13PRA}. 
Other experiments, where both interferometer arms were equally accelerated and, thus, the relative momentum in the interferometer remained unaffected, involve even higher numbers in photon transfer~\cite{Parker18science,Xu19Science}.

Large momentum transfer is especially of interest for increasing the sensitivity of Sagnac interferometers which scales with the enclosed area.
Compared to laser gyroscopes exhibiting resolutions of up to $10^{-11}~\mathrm{rad/(s\sqrt{Hz})}$~\cite{Schreiber13RevScI}, in matter-wave interferometers~\cite{Canuel06PRL,Stockton11PRL,Berg15PRL,Savoie18SciAdv,Wu07PRL,Moan20PRL} smaller areas suffice to reach high sensitivities .
However, forming the required loop size is still challenging and drives the dimensions of the apparatus.

Compared to previous approaches twin-lattice interferometry with BECs displays several advantages:
(i) It enables large and symmetric momentum transfer and, hence, large space-time areas in a simple and efficient way by combining double Bragg diffraction (DBD)~\cite{Ahlers16PRL,Hartmann20PRA} with symmetric Bloch oscillations (BO)~\cite{Peik97PRA}.
(ii) The symmetry of our geometry will suppress systematic effects such as diffraction phases, a common challenge in current implementations~\cite{Buechner03PRA,Plotkin18PRL,Parker18science}.
(iii) Combined with a delta-kick-collimated BEC source, we achieve unprecedented scalability marked by low atom and contrast loss.
According to our theoretical model, the interference contrast achieved with this method so far is solely limited by technical features of our device. 

Consequently, twin-lattice interferometry allows creating large Sagnac loops in a swift manner in compact devices outperforming today's approaches~\cite{Savoie18SciAdv,Moan20PRL,Pandey19Nat}.
Furthermore, the method brings within reach the momentum transfer of thousands of photon recoils and opens up exciting perspectives for many applications, in particular gravitational wave detectors such as MIGA~\cite{Canuel18SciRep} and ELGAR~\cite{ELGAR}.

\section*{Results}
\subsection*{Experimental setup}\vspace*{-14pt}  
The setup for our twin-lattice interferometer is shown in Fig.~\ref{fig:setup}. 
The twin lattice is formed by retroreflecting a laser beam featuring two frequencies with linear orthogonal polarization through a quarter-wave plate (see section~S1). The beam of 3.75~mm waist and up to 1.2~W power travels below and parallel to the horizontally aligned surface of an atom chip. 

The latter serves to generate BECs of up to $1.5\times10^4~^{87}$Rb atoms in the magnetic state~$F=2,m_F=2$ in a Ioffe-Pritchard trap with frequencies of ($43,344,343$)~Hz as detailed in~\cite{Zoest10science}. 
The BECs are released from the trap in free fall by switching off the magnetic field and are, after a free expansion of~5.4~ms, again exposed to the same field for~0.3~ms. 
In this way, they experience a delta-kick, which acts as a collimating lens and results in a narrow momentum distribution~\cite{Muentinga13PRL}, i.e. a residual expansion rate along the lattice of~$\sigma_{v}=0.18$~mm/s or~$0.03{\hbar}k$, corresponding to an effective 1D-temperature of 340~pK. 
The collimation is crucial for efficient manipulation of the BECs via DBD in combination with momentum transfer through BO~\cite{Szigeti12NJP}. 
Immediately after the delta-kick, an adiabatic rapid passage of 9~ms duration transfers 90-95\% of the BEC to the non-magnetic state $F=2,m_F=0$ by coupling the Zeeman levels with a chirped radio-frequency pulse. 
This state is separated from the others by temporarily applying a vertical magnetic field gradient as in Stern-Gerlach-type experiments.
Atoms are detected by absorption imaging with a CCD camera at a maximum observable free-fall time of~35.5~ms.

\subsection*{Interferometer sequence}
\begin{figure}[h]
\centering
\includegraphics[width=\columnwidth]{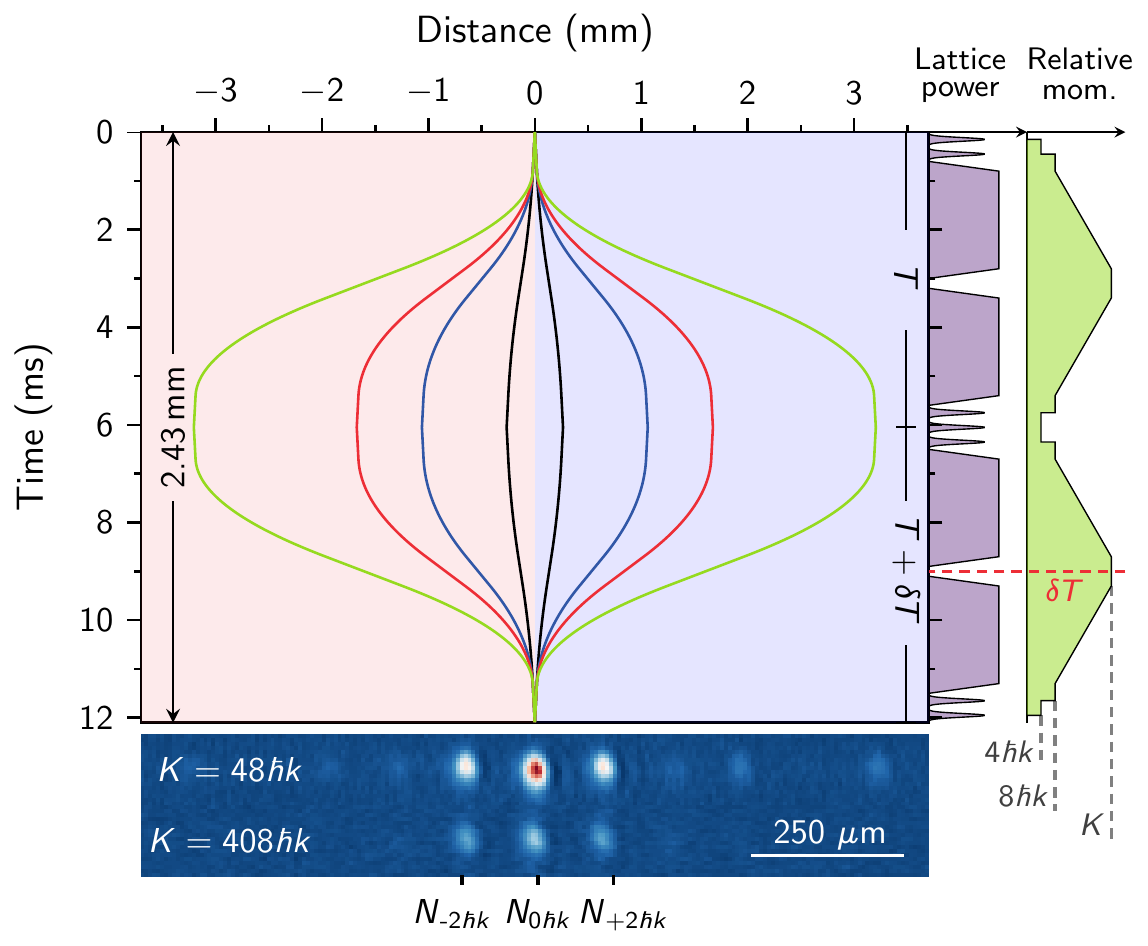}
\caption{\textbf{Twin-lattice scheme and sequence.} 
Space-time trajectories of the wave packets during the interferometer depicted for momentum transfers of~$K=(24,128,208,408){\hbar}k$ (black, blue, red, green) with distances given relative to the release point. 
Below: Absorption images of the interferometer output ports for~$K=48{\hbar}k$ and $408{\hbar}k$ after 35.5~ms of free fall.
Right: Temporal sequence of the twin-lattice laser power and the relative momentum~$K$ in the interferometer.
DBD is driven by Gaussian-shaped pulses of~$37.5~\mu$s width. 
BO for acceleration and deceleration are realized via a linear frequency ramp of 2~ms with~$200~\mu$s of loading and unloading time.
For contrast analysis, the free evolution time in the second half is modified by $\delta T$ with respect to the first half.
}
\label{fig:scheme}
\end{figure}

\vspace*{-14pt}  
The interferometer is generated by a sequence of DBD processes and BO in the twin lattice.
Fig.~\ref{fig:scheme} shows the resulting space-time trajectories of the wave packets exemplary for momentum transfers of $K=(24,128,208,408)\hbar k$ together with the corresponding temporal sequence of lattice power and change in relative momentum.

First, the twin lattice is exploited to induce two successive first-order DBD processes~\cite{Ahlers16PRL} so that the BEC is split into two wave packes separating with a mean momentum of~${\pm}4{\hbar}k$, respectively.
The light pulses driving DBD have a Gaussian-shaped temporal envelope of~37.5~$\mu$s width. Thanks to the use of delta-kick collimated BECs, we achieve a transfer efficiency of~98.8\% per recoil in these processes.
Hereafter, the BECs are loaded into the counterpropagating lattices.
During this process, the lattice intensity is linearly increased within 200~$\mu$s and the velocity adjusted to the comoving BECs~\cite{Denschlag02JPB}.
Each wave packet is accelerated by its copropagating lattice via BO during 2~ms and gains additional momentum of up to~$200{\hbar}k$.
For the release from the lattices, the intensity is lowered again linearly.
In this way, we obtain an efficiency of up to~99.93\% per recoil for BO in our interferometers.
In addition, we have realized a single beam splitter, where the wave packets have been accelerated up to a differential momentum of $K=1008\hbar k$ with a Bloch efficiency of~99.87\% per recoil.
Presently, the transferred number of photon recoils is limited solely by the free fall time and laser power available in our apparatus.

In the interferometer, the accelerated wave packets evolve freely for 200~$\mu$s, before their motion is first slowed down via BO to ${\pm}4{\hbar}k$, then inverted via successive DBD and again accelerated via BO. 
After a second free evolution of 200~$\mu$s, the velocities of the wave packets are reduced to ${\pm}4{\hbar}k$ to recombine them via the last DBD process resulting in three output ports with mean momenta of $\pm 2\hbar k$ and $0\hbar k$ showing interferences.
In total, the free evolution of the wave packets and their interaction with the light pulses amount to a duration of $2T=12.1$~ms.

The signal of our twin-lattice interferometer is defined as the normalized number of atoms detected in the output ports $p=(N_{+2{\hbar}k}+N_{-2{\hbar}k})/(N_{+2{\hbar}k}+N_{-2{\hbar}k}+N_{0{\hbar}k})$, where $N_{{\pm}2{\hbar}k}$ are the atom numbers with $\pm 2\hbar k$ and $N_{0{\hbar}k}$ with $0\hbar k$ momentum.
The absorption images at the bottom of Fig.~\ref{fig:scheme} show the output ports for two experimental realizations where the wave packets in the interferometer have had a momentum splitting of~$K=48{\hbar}k$ or~$408{\hbar}k$, respectively.

\subsection*{Contrast analysis}
\begin{figure}[htb]
\begin{center}
\includegraphics[width=\columnwidth]{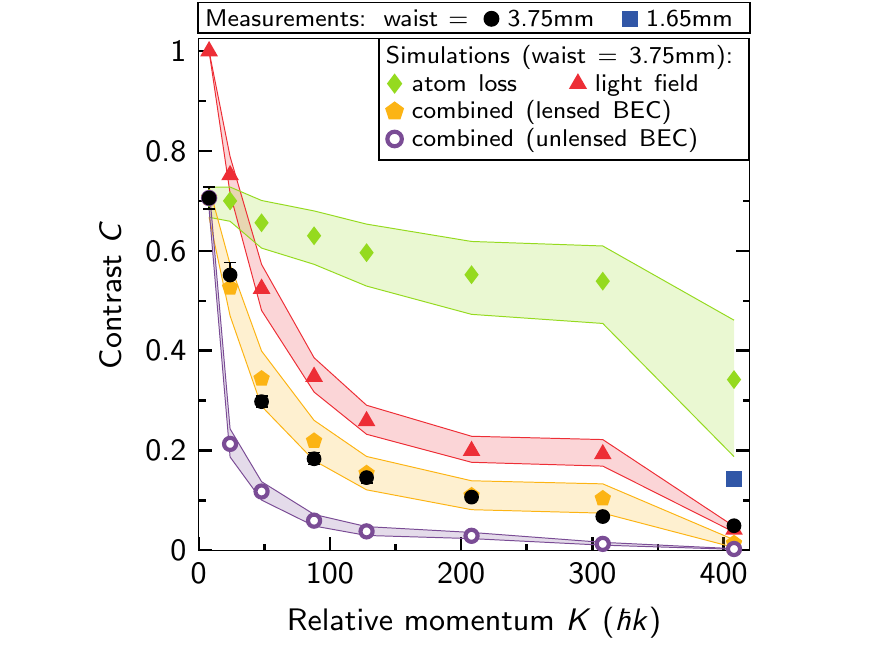}
\caption{\textbf{Contrast analysis.}
Experimental (black circles) and theoretical contrast $C$ of our twin-lattice interferometer in dependence of relative momentum~$K$ for a laser beam waist of 3.75~mm. 
We assume two effects dominantly contributing to the observed loss of contrast:
The first is atom loss due to non-adiabatic transitions~(green diamonds).
The second effect is a local inhomogeneous dipole force due to light field distortions~(red triangles).
Combining the simulation of both effects (orange pentagons) leads to a reasonable agreement with the experimental observations.
According to our model, a non-collimated BEC (open violet circles) would suffer more strongly from light field distortions and exhibit a significantly lower contrast due to its larger cloud size. 
Since we assumed equal atom losses for DBD and BO as for the collimated cloud, this presents an upper bound for the contrast. 
The shaded areas represent confidence intervals of the simulation, determined by atom number and lattice depth uncertainty.
The blue square shows a measurement, where the contrast has been improved by reducing the twin-lattice beam waist and, hence, spatial distortions of the twin-lattice beam due to diffraction at the atom chip.
}
\label{fig:contrast}
\end{center}
\end{figure}

\vspace*{-14pt}  
We evaluate the contrast of our interferometer statistically as described in detail in Materials and Methods. As illustrated in Fig.~\ref{fig:contrast} the measured interference contrast $C$ (black dots) clearly decreases with an increasing number of transferred momenta. 
Confirmed by simulations, we attribute the contrast loss to two main effects.

The first effect stems from atom losses arising during DBD and Bloch oscillations. On the one hand, atoms that have not been Bragg diffracted to the desired momentum states, and therefore have not performed BO, still give rise to an offset signal in the number of detected atoms. 
On the other hand, losses during BO remove atoms from the interferometer and reduce the number of coherent atoms. 
Given the limited free-fall time of the BECs in our apparatus, we had to enlarge the acceleration of BO to increase the relative momentum $K$ causing larger Landau-Zener losses.
Both loss mechanisms lead to a reduced contrast of about $C=0.35$ for $K=408\hbar k$ (green diamonds, see sections S2 and S3). 

The second cause for the contrast reduction is dephasing due to an imperfect Gaussian beam profile (see Fig.~S1 and section~S4 for more details).
Such perturbations result e.g. from light diffraction on an edge of the atom chip and induce spatially variable dipole forces along the two interferometer arms.
Different atomic trajectories according to the ensemble's spatial distribution in combination with path-dependent dipole forces lead to a spatially varying phase across the wave packets~\cite{Charriere12PRA,Roura14NJP,Bade18PRL} and give rise to a loss of contrast (red triangles). 
Combining the models for atom loss and dephasing processes (orange pentagons) enables us to match the observed contrast. We base our analysis of the beam profile on a model assuming a mere clipping of the twin lattice beams at one edge of the chip. Hereby, the relative magnitude of the intensity perturbations represents the only free fitting parameter. 
The results in Fig.~\ref{fig:contrast} have been obtained with a value equal to 9\% of the twin-lattice depth, which is compatible with the measured beam profile. 

In our simulation, the loss of contrast critically depends on the wave packet size (see also Fig.~S2).
The twin-lattice interferometer therefore benefits from the small spatial extent of our delta-kick collimated BEC, exhibiting Thomas-Fermi radii of $R=(32,39,30)~\mu$m at the end of the interferometer.
In the absence of delta-kick collimation these radii increase by a factor of three in $y$ and $z$-direction.
Since a larger ensemble samples over more light field distortions and, hence, suffers from stronger dephasing, it leads to a significant contrast reduction (open violet circles).
Our calculation even overestimates the contrast of the non-collimated BEC, since it is performed assuming the same efficiencies for DBD and BO as in the case of delta-kick collimated BECs. 

To experimentally verify the impact of the spatial profile of the light field on the interferometric contrast we have reduced the beam diameter by roughly a factor of two compared to our original setup.
The smaller diameter lowers light field distortions caused by diffraction at the atom chip and other apertures.
In this way, we are able to triple the measured contrast in our largest interferometer ($K=408{\hbar}k$) to $C=0.14$ (blue square).

\section*{Discussion}
\subsection*{Scalibility}\vspace*{-14pt}
We have employed twin-lattice interferometry for large symmetric momentum transfer and observed spatial coherence up to a maximum splitting of~$K=408{\hbar}k$, the largest momentum separation in an interferometer reported so far.
In comparison to previous results, our method is not as strongly limited by atom loss as the benchmark experiment in~\cite{Chiow11PRL}. 
Moreover, diffraction phases should be greatly suppressed as in~\cite{Pagel19arxiv}, however, the inertial noise in our current setup prevents us from confirming this experimentally.
Indeed, for ideal mode overlap, our twin-lattice interferometer should be by construction not susceptible to light shifts that were reported in~\cite{Clade09PRL}.

Our symmetric geometry also significantly relaxes laser power requirements compared to an asymmetric scheme. 
For the same momentum transfer, the acceleration of only a single interferometer arm would require larger lattice depths causing not only higher atom losses due to spontaneous scattering but also an even lower contrast due to the light field distortions.

Compared to former symmetric schemes, we significantly exceed the contrast observed in the double Raman interferometer~\cite{Leveque09PRL} and our own double Bragg interferometer~\cite{Ahlers16PRL}.
Our experimental studies, as well as theoretical simulations, reveal that only technical reasons, in particular spatial distortions of the twin lattice, limit the current beam splitter efficiency and generally the scaling of our method.

In fact, our theoretical model for Gaussian-shaped beams and experimental results show that in the ideal case of an undisturbed lattice, neither the Landau-Zener losses nor the differential dipole force arising due to the Gaussian waist will be critical.
For a twin lattice featuring spatial intensity fluctuations in the order of~0.5\% of $V_0$ the model predicts a contrast of more than 90\% at~$K=408{\hbar}k$.
An additional absolute Stark shift compensation may help to reduce the effect of residual local inhomogeneities and maintain a high contrast by adding a light field of opposite detuning~\cite{Kovachy15Nature,Parker18science}.

\subsection*{Current and future sensor performance}\vspace*{-14pt}
Twin-lattice interferometry allows to efficiently create large Sagnac areas in comparably short time and space.
To evaluate the level of miniaturization and to compare the geometry obtained with our method to other sensors, we introduce a compactness factor $(L\cdot \tau)^{-1}$, which is the inverse product of interferometer time $\tau$ and baseline $L$.

\begin{figure}[ht]
\begin{center}
\includegraphics[width=\linewidth]{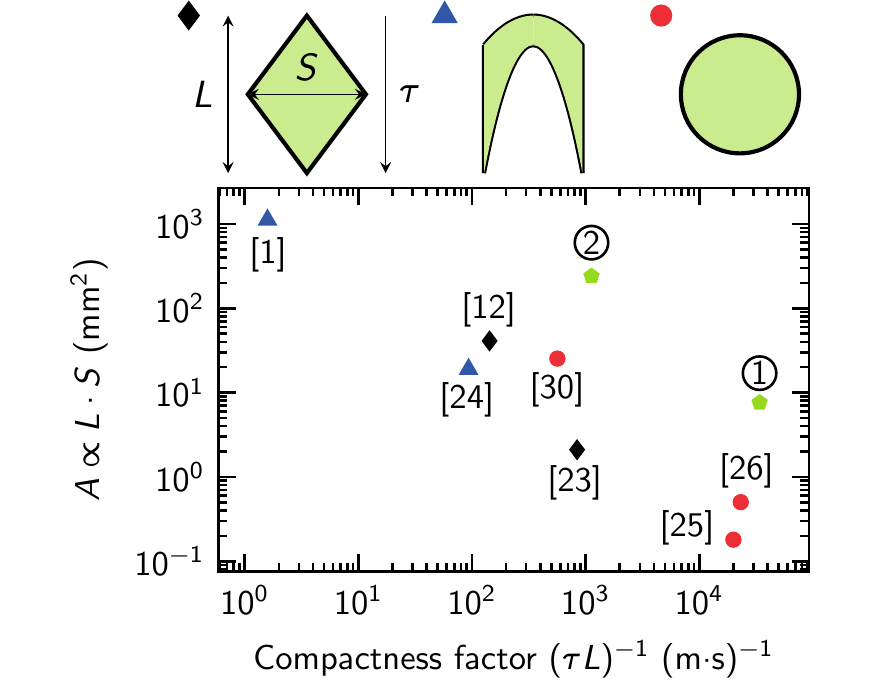}
\caption{\textbf{Sensor comparison.}
Comparison of the effective area $A$ enclosed by the interferometer arms as a function of the compactness factor $(\tau \cdot L)^{-1}$, the inverse of the product of interferometer duration $\tau$ and baseline $L$.
In principle, the area increases linearly with the maximum separation $S$ and the baseline $L$.
For high compactness and large Sagnac areas it is advantageous to reach large separations $S$ in short times $\tau$ at a balanced baseline $L$.
The presented experiments employ either Mach-Zehnder type topologies (black diamonds)~\cite{Canuel06PRL,Berg15PRL}, butterfly geometries (blue triangles) \cite{Stockton11PRL,Savoie18SciAdv} or ring-shaped guides (red circles) \cite{Pandey19Nat,Moan20PRL,Wu07PRL}. 
Green pentagons depict our twin-lattice interferometer in the current \raisebox{-1.5pt}{\Large\textcircled{\raisebox{0.9pt} {\small 1}}} ($\tau=12.1~$ms, $K=408\hbar k$) and a future \raisebox{-1.5pt}{\Large\textcircled{\raisebox{0.9pt} {\small 2}}} version with improved interferometer parameters ($\tau=48.4~$ms, $K=808\hbar k$).  }
\label{fig:gyroscopes}
\end{center}
\end{figure}

Fig.~\ref{fig:gyroscopes} shows the shape and size of areas employed in different gyroscopes as a function of their compactness factor. 
Typical butterfly-geometries~\cite{Stockton11PRL,Savoie18SciAdv} achieve high sensitivities but feature only little compactness.
In our twin-lattice interferometer ($K=408\hbar k$) with a total duration of $\tau =2T =12.1~$ms we enclose an area as large as $A=7.6~$mm$^2$ at a baseline of merely $L=2.43~$mm. 
Overcoming current technical limitations, the realization of an interferometer with $\tau=48.4$~ms and a momentum separation of $K=808\hbar k$ at a contrast of $C=0.5$ seems feasible.
This would lead to an increased area of $A=240$~mm$^2$ and a baseline of $L=18.3$~mm.
In combination with an advanced atom source~\cite{Rudolph15NJP} providing BECs of $N=10^5$ atoms at a 1~Hz rate, such a device would feature shot-noise limited sensitivities towards rotations and accelerations of $8\cdot 10^{-9}\,$rad/s and $1.6\cdot 10^{-9}\,$m/s$^2$ per cycle.
Hence, instead of requiring meters of baseline, this twin-lattice interferometer would fit into a volume of less than 1~cm$^3$. 
In terms of miniaturization it is comparable to guided structures~\cite{Pandey19Nat,Moan20PRL,Wu07PRL}.

In conclusion, the symmetric nature and unprecedented scalability make twin-lattice atom interferometry an excellent candidate for applications requiring large space-time areas.
Moreover, the demonstrated efficiencies recommend to combine twin-lattice interferometry with BECs featuring non-classical correlations~\cite{Esteve08Nature,Lange18Science}. 
Besides gyroscopes~\cite{Stockton11PRL,Savoie18SciAdv}, our method is suitable for enhanced quantum tilt meters~\cite{Ahlers16PRL,Xu17PRL}, gradiometers~\cite{Asenbaum17PRL}, and~$h/m$ measurements~\cite{Parker18science}.
Last but not least, twin-lattice interferometry should open up the path to devices such as MIGA~\cite{Canuel18SciRep} and ELGAR~\cite{ELGAR} employing BECs with relative momenta of one thousand photon recoils as required for terrestrial detectors of infrasound gravitational waves~\cite{Schubert19}.

\begin{figure}[h]
\begin{center}
\includegraphics[width=\linewidth]{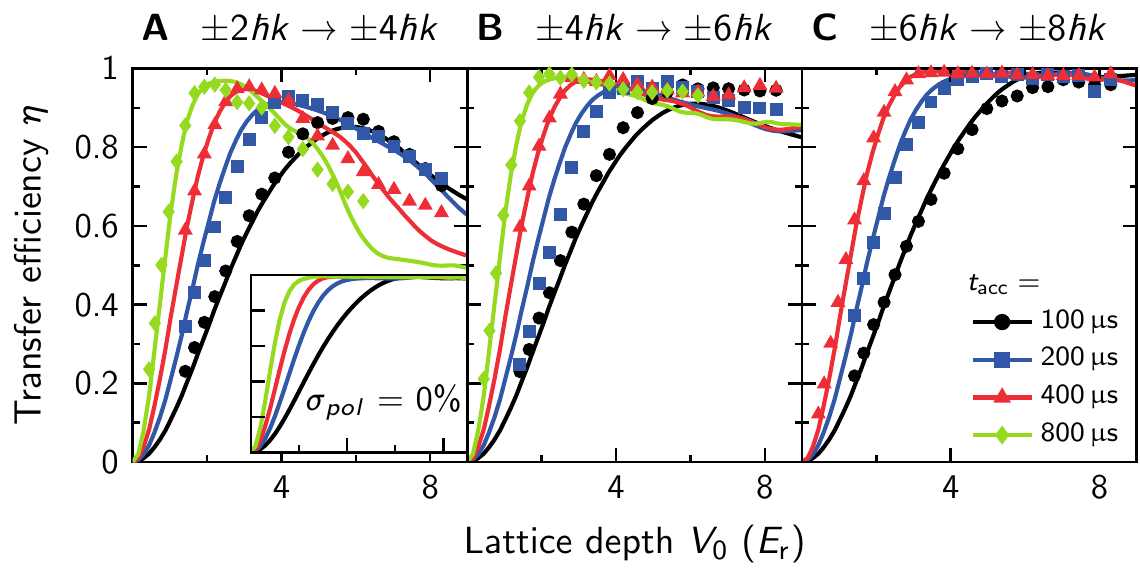}
\caption{\textbf{Bloch transfer efficiency in the twin lattice.} Efficiency of one BO exerted by our twin lattice in dependence of the initial state. 
A superposition of BECs with momenta (\textbf{A})~${\pm}2{\hbar}k$,(\textbf{B})~${\pm}4{\hbar}k$ or (\textbf{C})~${\pm}6{\hbar}k$ created via first-order DBD and sequential Bragg pulses is accelerated via BO acquiring an additional momentum of~${\pm}2{\hbar}k$. 
The efficiency is recorded for different times~$t_\mathrm{acc}$ and in dependence of the lattice depth~$V_\mathrm{0}$ in units of recoil energy $E_r = \hbar^2 k^2 /(2m) $.
In all cases, a larger~$t_\mathrm{acc}$ corresponds to a lower acceleration and reduces non-adiabatic losses as expected from Landau-Zener theory.
Our experimental results (symbols) are well reproduced by theoretical simulations (lines) assuming an additional standing lattice of depth $0.37 V_\mathrm{0}$ arising from polarization imperfections. The losses caused by the latter decrease with a larger initial splitting~$K$ and from (\textbf{A}) to (\textbf{C}) as the detuning from the standing lattice increases. The inset shows the theoretical efficiency for ideal polarization, i.e. a perfect twin lattice, of the transfer ${\pm}2{\hbar}k\rightarrow{\pm}4{\hbar}k$.}
\label{fig:landauzener}
\end{center}
\end{figure}

\section*{Materials and Methods}

\subsection*{Twin-lattice laser system}\vspace*{-14pt} 
For interferometry, we employ a frequency-doubled fiber laser system~\cite{Abend16PRL}(NKT Photonics Koheras Boostik and Toptica Photonics SHG pro) 100~GHz blue-detuned from the $^{87}$Rb D$_2$ line to reduce spontaneous emission.
To create the twin lattice we use two light fields with orthogonal linear polarizations, which are merged to a single beam on a polarizing beam splitter. Their frequencies and amplitudes are controlled independently by acousto-optical modulators (AOM, AA Opto Electronics, MT80-A1.5-IR). 
The light is guided to the atoms via a single-mode fiber and collimated to a Gaussian-shaped beam of 3.75~mm waist and up to 1.2~W power (Thorlabs, F810APC-780). 
By retroreflecting the beam through a quarter-wave plate the twin lattice is formed, i.e. two lattices counterpropagating along the horizontal direction with orthogonal polarization to avoid standing waves.
The lattices can be simultaneously accelerated in opposite direction by chirping their frequency difference with the AOMs.

The radio frequency signal for the AOMs is generated by two different devices. For BO, we employ a self-built device which is able to drive linear amplitude and frequency ramps at a sample rate of 250\,kHz. For DBD, we use a pulse generator (PulseBlasterDDS-II-300-AWG) to form a Gaussian-shaped temporal intensity envelope.

\subsection*{Bloch transfer efficiency in the twin lattice}\vspace*{-14pt}  
High efficiencies in the transfer of photon recoils are crucial for generating symmetric interferometers employing large momentum separation.

\newpage
\onecolumngrid
\begin{figure*}[htb]
\begin{center}
\includegraphics[width=\linewidth]{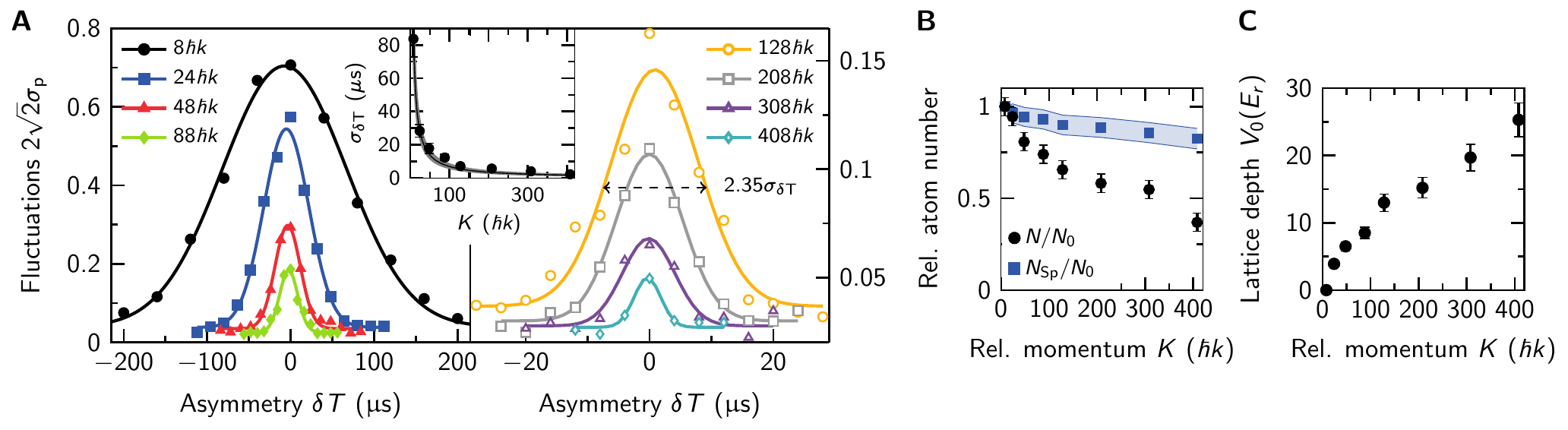}
\caption{\textbf{Interferometer analysis.}
(\textbf{A}) Measured atom number fluctuations~$2\sqrt{2}\sigma_p$ of the twin-lattice interferometer in dependence of the spatial overlap of the wave packets at the last DBD pulse for increasing relative momenta~$K$ and corresponding Gaussian fits.
The contrast is proportional to the standard deviation~$\sigma_p$ of the normalized atom number~$p$ at the output port.
We modify the spatial overlap by varying the duration of the second free-evolution time by~${\delta}T$ relative to the first one.
The displacement of the two wave packets varies with~$\delta{T}K$.
Inset: The width~$\sigma_{{\delta}T}$ of the envelopes decays with the inverse of the relative momentum~$K$ for a constant spatial coherence length. The line represents the theoretically calculated value $\sigma_{{\delta}T}= \hbar / (K \sigma_v)$ for $\sigma_v = (0.18 \pm 0.03)$~mm/s. 
(\textbf{B})~Relative atom number~$N$/$N_0$ measured in the output ports normalized to the DBD interferometer~($K=8{\hbar}k$) (black circles) and calculated spontaneous emission decay~$N_{Sp}$/$N_0$ with confidence intervals (blue squares).
(\textbf{C})~Lattice depth~$V_0$ applied during acceleration and deceleration with BO. Atom number and lattice depth serve as input parameters for our contrast simulation.
}
\label{fig:envelopes}
\end{center}
\end{figure*}
\twocolumngrid

Since, in case of a twin lattice, one lattice might disturb the effect of the other, we compare the efficiency of BO for different initial momentum separations of the BEC. Ideally, they should be equal to the ones of the individual lattices.
Fig.~\ref{fig:landauzener} illustrates non-adiabatic losses during BO in a twin lattice.
We first prepare a superposition of symmetric momentum states~${\pm}2{\hbar}k$~(A), ${\pm}4{\hbar}k$~(B) or ${\pm}6{\hbar}k$~(C) by performing first-order DBD and a respective number of sequential first-order Bragg pulses.
Subsequently, an additional momentum of~${\pm}2{\hbar}k$ is imparted via BO and the number of atoms in the final momentum states is measured.
In particular, we compare the experimentally achieved (symbols) and the matching theoretically simulated efficiency (lines) for different durations $t_\mathrm{acc}=(100,200,400,800)~\mu$s implying different accelerations in dependence of the lattice depth~$V_\mathrm{0}$. 
In general, the transfer efficiency decreases for an increasing acceleration. Indeed, slower accelerations lower the rate of non-adiabatic interband transitions, and, hence, reduce atom losses in the final momentum state as expected from Landau-Zener theory~\cite{Peik97PRA}.
Figs.~\ref{fig:landauzener}A-C clearly show the importance of the initial momentum separation prepared by DBD for an efficient transfer by the subsequent BO in our setup. While the efficiency for larger initial separations equals almost a hundred percent, it is lower for smaller separations and also exhibits a pronounced optimum.

To explain the observed atom losses associated with the initial state, our simulations, based on a 1D-reduced Gross-Pitaevskii model~\cite{Salasnich02PRA}, had to take into account imperfections of our optical setup creating the twin lattice (see section~S1 for more details).
Neither the Landau-Zener theory describing an ideal single lattice nor our simulations of an ideal twin lattice, as shown in the inset (Fig.~\ref{fig:landauzener}A) for the process~${\pm}2{\hbar}k\rightarrow{\pm}4{\hbar}k$, could reproduce the experimental data.
Indeed, these losses result from an imperfect polarization of the light fields giving rise to unwanted standing waves. 
Adding a standing lattice of~$0.37 V_\mathrm{0}$ depth, which is compatible with our experimental setup, the theoretical curves agree with the experimental data.
Fortunately, the losses can be overcome by increasing the initial momentum splitting of the BEC in our setup, which corresponds to a larger detuning to standing waves.
In consequence, we start our interferometers with a beam splitter creating a superposition of~${\pm}4{\hbar}k$ (as in Fig.~\ref{fig:landauzener}B), being the best trade-off between the losses caused by parasitic standing waves and the lower efficiency of Bragg processes.

Fig.~\ref{fig:scheme} depicts the temporal sequence regarding laser power and relative momentum for the combination of DBD with BO in our twin lattice.
For acceleration and deceleration of the BECs via BO the frequency difference of the twin lattice is linearly ramped up or down, respectively.
For relative momenta of $K = (24, 48, 88)\hbar k$ a single linear frequency ramp of 2~ms is employed.
If the momentum transfer $K$ in the interferometer is larger, the frequency ramp is split into two parts. For acceleration (deceleration) to relative momenta of $K = (128,208,308,408)\hbar k$ a momentum of $12\hbar k$ is transferred during the first (final) $(0.5,0.4,0.3,0.3)~$ms and the rest in the remaining $(1.5, 1.6, 1.7, 1.7)~$ms.

\subsection*{Evaluation of spatial coherence}\vspace*{-14pt}  
We analyze the spatial coherence of our twin-lattice interferometers statistically by measuring the interference amplitude in dependence of a spatial separation of the wave packet trajectories at the moment of the last DBD pulse~\cite{Kovachy15Nature}.
The displacement between the wave packets is modified by varying the free evolution time in the second half of the interferometer by~$\pm{\delta}T$ as indicated in Fig.~\ref{fig:scheme}. 
The statistical analysis allows evaluating the spatial coherence in the presence of large vibrational noise.
In our setup seismic measurements reveal noise as large as~$10^{-2}$~m/s$^2/\sqrt{\mathrm{Hz}}$. 
Even in the case of interferometers involving momentum states as low as~${\pm}4{\hbar}k$ and a duration as short as~$2T=12.1$~ms this leads to a phase noise significantly larger than $2\pi$.
Hence, we assume that the population of the interferometer ports results from a random phase, i.e. white phase noise.

Ideally, the displacement of the trajectories at the final DBD pulse is proportional to the time difference~$\delta{T}$ and the relative velocity~$K/m_\mathrm{Rb}$.
The dependence of the contrast on the displacement is described by a Gaussian bell-shaped curve~$2\sqrt{2}\sigma_p(\delta{T})\exp(-(K/\hbar)^2\sigma_{v}^2{\delta}T^2/2)$~\cite{Kovachy15Nature}, $\sigma_p$ being the standard deviation of the normalized population~$p$ and~$\sigma_{v}$ the expansion rate of the BEC along the lattice.
To calculate $\sigma_p$ we take 40 data points for each ~$\delta{T}$ (see Fig.~S3).

Fig.~\ref{fig:envelopes}A shows the experimentally determined value of $\sigma_p$ depending on~$\delta{T}$ for different relative momenta~$K$ as well as the corresponding Gaussian fits to the data.
The latter allow us to determine the maximum contrast $C$ and the distribution's width~$\sigma_{{\delta}T}$.
For a given spatial coherence length, determined by $\sigma_{v}$, the width~$\sigma_{{\delta}T}=\hbar/(K\sigma_{v})$ ideally reduces with the inverse of the momentum separation~$K$ (inset in Fig.~\ref{fig:envelopes}A).
This agrees well with our observations and indicates that the coherent manipulation of the BEC is not reducing the coherence length of the atomic ensemble contributing to the interference signal.

\subsection*{Input parameters for contrast simulation}\vspace*{-14pt}  
In our case, the experimentally determined values of the lattice depth~$V_0$ during BO and the total atom number~$N$ detected in the output ports serve as input parameters for our simulation of the interference contrast in the twin lattice (sections S3 and S4).
We measure $N$ relative to $N_0 \equiv N(8\hbar k)$ obtained in an interferometer exclusively based on DBD (Fig.~\ref{fig:envelopes}B).
The fraction $N/N_0$ declines with relative momentum $K$ to about 35\% for $K=408 \hbar k$ since non-adiabatic and spontaneous emission losses rise with increasing accelerations that require deeper lattices (Fig.~\ref{fig:envelopes}C).
We calculate $N_\mathrm{Sp}$, namely the atom number $N_0$ diminished by spontaneous emission losses in a twin lattice of depth $V_\mathrm{0}$ (see section~S2 for details).
For example, in case of the~$K=408{\hbar}k$ interferometer the required lattice depth $V_\mathrm{0}=25.3 E_r$ leads to a spontaneous emission rate of 22~s$^{-1}$ causing an atom loss of $1 - N_\mathrm{Sp}/N_0 = 18\%$. 
In our model, spontaneous scattering equally affects atoms contributing to the signal and to the offset and, hence, does not influence the contrast.

The confidence intervals result from a 10\% error in $V_0$ and an absolute error of 0.05 regarding $N/N_0$.

\section*{References and notes}

\vspace*{14pt}  
\section*{Acknowledgments} 
\small
{\bfseries General:}
We thank E. Giese, A. Friedrich, J. Jenewein, A. Roura, Z. Pagel, M. Jaffe, P. Haslinger and H. M\"uller for fruitful discussions.

{\bfseries Funding:}
The presented work is supported by the German Research Foundation (DFG) under Germany's Excellence Strategy (EXC-2123 QuantumFrontiers Grant no. 390837967) within reasearch units B02 and B05 and through the CRC 1227 (DQ-mat) within Projects No. A05, No. B07 and No. B09, as well as through the CRC 1128 (geo-Q) within the Projects No. A01 and No. A02.
We also acknowledge support from the QUEST-LFS, the German Space Agency (DLR) with funds provided by the Federal Ministry of Economic Affairs and Energy (BMWi) due to an enactment of the German Bundestag under Grant No. DLR 50WM1952 and 50WM1955 (QUANTUS-V-Fallturm), 50WM1642 (PRIMUS-III), 50WM1861 (CAL), 50WP1700 (BECCAL), 50RK1957 (QGYRO), the Verein Deutscher Ingenieure (VDI) with funds provided by the Federal Ministry of Education and Research (BMBF) under Grant No. VDI 13N14838 (TAIOL).
We acknowledge financial support from “Nieders{\"a}chsisches Vorab” through the ``Quantum- and Nano-Metrology (QUANOMET)'' initiative within the project QT3 and through “F{\"o}rderung von Wissenschaft und Technik in Forschung und Lehre“ for the initial funding of research in the new DLR-SI Institute.

{\bfseries Author contributions:}
All authors contributed to the scientific discussion and approved the final manuscript. M.~Gebbe and J.-N.~Siem{\ss} contributed equally to this work. M.~Gebbe, M.~Gersemann and S.~Abend performed the experiments and analyzed the data. J.-N.~Siem{\ss} and N. Gaaloul developed the theoretical model. M.~Gebbe, J.-N.~Siem{\ss}, S.~Abend and E.M.~Rasel prepared the manuscript. K.~Hammerer, C.~L\"ammerzahl and E.M.~Rasel supervised the project.

{\bfseries Competing interests:}
All authors declare that they have no competing financial and nonfinancial interests. 

{\bfseries Data and materials availability:}
All data generated or analyzed during this study are included in this published article or are available from the corresponding author on reasonable request.

\normalsize

\clearpage
\setcounter{figure}{0}  

\onecolumngrid

\newgeometry{top=2.2cm,bottom=2.2cm,left=2.2cm,right=2.2cm}

\begin{center}
  \Large Supplementary Materials for \\\vspace{0.5cm}
\Large Twin-lattice atom interferometry   
\end{center}

\vspace{0.5cm}

\baselineskip18pt

\section*{Section S1. Simulation of twin-lattice beam splitter efficiency}
The twin lattice is formed by the electric field~$\vec{E}(y,t)$ propagating along the horizontal axis, here in $y$-direction. As depicted in Fig.~\ref{fig:setup} it comprises two frequencies~$\omega_1$ and~$\omega_2$ 
\begin{align} \label{eq:electric_field}
\vec{E}(y,t) = \left[  E_\mathrm{1}\vec{\epsilon}_1  \cos{\left(k_1y - \omega_1 t\right)} + E_\mathrm{2} \vec{\epsilon}_2 \cos{\left(k_2y - \omega_2 t\right)}\right] \; ,
\end{align} 
and passes a quarter-wave plate before and after retroreflection.
Ideally, the twin lattice features equal field amplitudes of both frequency components~$E_0 \equiv E_1 = E_2$. We define~$\Delta \omega  = \omega_1 - \omega_2$ and note that~$k \equiv k_1\approx k_2$. Within the rotating wave approximation, the atom-light interaction gives rise to the spatially modulated dipole potential with depth $V_0$: 
\begin{align} \label{eq:latticepot}
V(y,t) =V_0 \left[ \vphantom{\frac{1}{2}}
\cos^2{\left(k y- \Delta \omega t/2 \right)} + \cos^2{\left(k y+ \Delta \omega t/2 \right)}  + \frac{\sigma_{\mathrm{pol}}}{(1- \sigma_\mathrm{pol})} \big[ 2 \cos{\left(k y +\Delta \omega t /2 \right)}\cos{\left(k y- \Delta \omega t /2\right)}\big]\right]
\end{align} 
In addition to the twin lattice (given by the terms~$\displaystyle \cos^2{\left(k y\pm \Delta \omega t/2 \right)}$) Eq.~(\ref{eq:latticepot}) features an interference term resulting from polarization imperfections, acting as a standing lattice ($\propto \cos^2(k y)$) in terms of momentum transfer. Its magnitude depends on the scalar product of the polarization vectors~$\vec{\epsilon}_1$~and~$\vec{\epsilon}_2 $:~$\sigma_\mathrm{pol} \equiv \left|{ \vec{\epsilon}_1 \cdot \vec{\epsilon}_2}\right|/2 \leq 0.5$.

We adapt a time-dependent Gross-Pitaevskii (GP) model~\cite{Salasnich02PRA} using the optical potential in Eq.~(\ref{eq:latticepot}) to calculate the efficiency of the momentum transfer depicted in Fig.~\ref{fig:landauzener}.
The initial atomic state for our simulations is obtained by calibrating 3D numerical GP simulations to the experimentally observed free evolution of the atomic wave packet including the release from the magnetic trap and delta-kick collimation. 
In our simulations, double Bragg diffraction (DBD) realizing the initial relative momentum $K_\mathrm{DBD}$ is described by creating an ideal superposition of counter propagating wave packets in position space. The atomic state is multiplied with the phase factors $\frac{1}{\sqrt{2}}e^{\pm i \varphi_{\text{DBD}}}$ (including normalization) providing the constituents of the superposition, where $\varphi_{\text{DBD}} = K_\mathrm{DBD}\, y/2\hbar$. 
Since our focus here is on the Bloch oscillation efficiency, we idealize the double Bragg interaction neglecting effects such as velocity selectivity or off-resonant couplings~\cite{Hartmann20PRA}. %

By changing the relative orientation of the polarization vectors we vary the strength of the undesired contributions to the dipole potential according to Eq.~(\ref{eq:latticepot}) which vanish in case of orthogonality. 
The theoretical curves presented in Fig.~\ref{fig:landauzener} have been obtained with~$\sigma_\mathrm{pol}  = 0.2688$ for all data sets corresponding to a standing wave depth of $0.37~V_0$ (cf. second line in Eq.~\ref{eq:latticepot}) which is a plausible assumption for the experimental setup.
The experimental values for $V_0$ in Fig.~\ref{fig:landauzener} have been calibrated via the Landau-Zener formalism.

\section*{Section S2. Spontaneous emission rate for atoms in moving optical lattices}
The spontaneous emission rate of an atomic transition of frequency~$\omega_A$ interacting with the twin-lattice potential in Eq.~(\ref{eq:latticepot}) can in general be expressed as~\cite{Clade06PRA} 
\begin{align} \label{eq:spon_emission}
    P = \frac{\Gamma}{|\Delta|} ~ \frac{\left<V(y,t)\right>}{\hbar}\;,
\end{align}
where~$\Gamma$ is the natural linewidth of the transition and~$\Delta \equiv \omega_{\mathrm{L}}-\omega_{\mathrm{A}} $ the detuning of the laser frequency~$\omega_{\mathrm{L}}$ from resonance. 
The light creating the twin lattice is blue detuned ($\Delta > 0$) and we evaluate the spontaneous emission rate in the rest frame of the atomic wave packets. 
We do so by substituting~$y\rightarrow y+\omega_\mathrm{D} t / k$ in $V(y,t)$, where~$\omega_D = k~v_\mathrm{BEC}$ depends on the atomic mean velocity~$v_{\mathrm{BEC}}$.\\
Without loss of generality, we consider wave packets copropagating with the lattice traveling in negative $y$-direction, i.e. $\omega_D = + \Delta \omega/ 2$, giving us the potential in the wave packet frame:
\begin{align} \label{eq:latticepotCOM}
 V(y,t)  = V_0  \left[\cos^2{\left(k y \right)} + \cos^2{\left(k y+ \Delta \omega t \right)} + \frac{\sigma_{\mathrm{pol}}}{(1- \sigma_\mathrm{pol})} \big[ 2 \cos{\left(k y +\Delta \omega t \right)}\cos{\left(k y\right)}\big]\right] 
\end{align} 
\\
In order to calculate the average spontaneous emission rate $P$ we consider the mean intensity an atom experiences during interaction with the lattice.
We evaluate the contribution of the first term in Eq.~\eqref{eq:latticepotCOM} associated with the copropagating lattice by considering that the atomic wave function is largely overlapping with the nodes of the repulsive lattice potential. 
As detailed in~\cite{Clade06PRA}, one can assume the atom to be well described by the lowest Bloch state of the comoving lattice to evaluate the average potential $\left<\cos^2{\left(k y \right)}\right> = \frac{1}{2} \sqrt{ E_\mathrm{r}/V_0}$, where $E_\mathrm{r} = \hbar^2 k^2/(2m)$ is the recoil energy. 
Given that the frequency within the cosine arguments $\Delta \omega$ is typically much greater than the rate~$ \frac{\Gamma}{|\Delta|} \frac{V_0 }{\hbar}$ in Eq.~\eqref{eq:spon_emission} we can take the temporal average of the other terms $\left<\cos^2{\left(k y + \Delta \omega t \right)}\right> \approx \frac{1}{2}$ and $\left<\cos{\left(k y + \Delta \omega t \right)} \cos{(ky)}\right> \approx 0$, respectively.
Inserting these results into Eq.~(\ref{eq:spon_emission}) leads us to the total emission rate for an atom that copropagates with one of the twins
\begin{align}
    P = \frac{\Gamma}{|\Delta|} ~ \frac{V_0}{\hbar}
 \left[\frac{1}{2\sqrt{V_\mathrm{0}/E_\mathrm{r}}} + \frac{1}{2}\right] \; . 
\end{align}

After a total duration~$\tau_\mathrm{twin}$ of the interaction with the twin lattice, the atom number~$N_\mathrm{Sp}$ thus decreases with rate~$P$ as
\begin{align}
   N_\mathrm{Sp} = N_0 \cdot e^{ - P\cdot \tau_\mathrm{twin}} \; ,
\end{align}
where~$N_0$ is the atom number measured in the output ports of the interferometer solely based on double Bragg diffraction ($K=8\hbar k$). Figure~6B displays the ratio~$N_\mathrm{Sp}/N_0$ in dependence of the relative momentum $K$. As~$\tau_\mathrm{twin}$ is identical for all our interferometer sequences featuring Bloch oscillations, the spontaneous emission rate~$P$ only depends on the lattice depth~$V_\mathrm{0}$ (Fig.~\ref{fig:envelopes}C).

\section*{Section S3. Contrast reduction due to inefficiency of the momentum transfer}
Atom losses during Bragg diffraction and Bloch oscillations degrade the contrast by reducing the coherent fraction of atoms which contribute to the interferometer signal. To estimate the contrast loss, we make the following assumptions to simplify our calculations:
(i) A finite double Bragg beam splitting fidelity leads to an offset $N_\mathrm{OS}$ in the output ports and, thus, to a reduction of the interference amplitude. 
(ii) Spontaneous scattering equally affects the coherent fraction of atoms as well as the offset in the interferometer ports and, hence, does not lead to a contrast decay in our case. 
Since we are using delta-kick collimated BECs with a momentum width far below the photon recoil, the vast majority of spontaneously scattered atoms is separated from the output ports, anyway, and not counted by our spatial detection system. 
(iii) Landau-Zener losses during Bloch oscillations remove atoms from the interferometer and, therefore, lead to loss of contrast.

The contrast can be expressed by the maximum~$p_{\mathrm{max}}$ and minimum~$p_{\mathrm{min}}$ normalized number of atoms detected in the interferometer ports:
\begin{align}\label{eq:contrast_general}
    C = 2\sqrt{2} \sigma_p \approx \frac{p_{\mathrm{max}} - p_{\mathrm{\mathrm{min}}}}{p_{\mathrm{\mathrm{max}}} + p_{\mathrm{\mathrm{min}}}},
\end{align}
where $\sigma_p$ is the standard deviation of the normalized population~$p$ defined in the main text.
We attribute the contrast of our double Bragg interferometer, $C(K = 8\hbar k) = 0.7059 \pm 0.022$, solely to the finite efficiency of the double Bragg beam splitting processes. Due to a similar fidelity for first and sequential Bragg diffraction the offset $N_\mathrm{OS}$ is assumed to be equally distributed between the inner ($N_{0\hbar k}$) and the outer ports ($N_{\pm 2\hbar k}$).
We write the atom number in the output ports $N$ as the sum of the signal $N_\mathrm{Sig}$ and the offset $N_\mathrm{OS}$, $N(K) = N_\mathrm{Sig}(K) + N_\mathrm{OS}(K)$. 

Our contrast model hinges on the following assumptions:
(i) The double Bragg processes are elements of all our interferometers and, hence, $N_\mathrm{OS}$ is identical for all~$K$. The extreme values for $p$ thus equal
\begin{align*}
   p_{\mathrm{min}} &= \frac{N_\mathrm{OS}(K) /2 }{N(K)}\;,\\
   p_{\mathrm{max}} &= \frac{N_\mathrm{Sig}(K) + N_\mathrm{OS}(K) /2}{N(K)} \;.
\end{align*}
Inserting these into Eq.~(\ref{eq:contrast_general}) leads to
\begin{align}\label{eq:contrast_fractions}
    C(K) = \frac{N_\mathrm{Sig}(K)}{N_\mathrm{OS}(K) + N_\mathrm{Sig}(K)}\;.
\end{align}
Since we assume an equal spontaneous emission rate for all momentum classes, $N_\mathrm{OS}$ decreases with rate $P$ in the presence of Bloch oscillations and can be written as
\begin{align*}
    N_\mathrm{OS}(K)=N_\mathrm{OS}(8\hbar k) \cdot e^{ - P(K) \cdot \tau_\mathrm{twin}}  = (1-C(8\hbar k)) N_0 \cdot e^{ - P(K) \cdot \tau_\mathrm{twin}} = (1-C(8\hbar k)) N_\mathrm{Sp}(K) \;.
\end{align*}

(ii) Acceleration of the atoms with Bloch oscillations lowers the signal~$N_\mathrm{Sig}(K)$ due to non-adiabatic transitions. The offset $N_\mathrm{OS}$ is only affected by spontaneous emission, but not by non-adiabatic losses, since offset atoms do not perform Bloch oscillations and Landau-Zener losses are expected to be spatially well separated from the output ports. 
A decrease in~$N_\mathrm{Sig}(K)$ therefore reduces the contrast according to Eq. \eqref{eq:contrast_fractions}. 

Combining these assumptions we express the contrast in Eq.~\eqref{eq:contrast_fractions} as a function of $K$ that requires as input parameters the experimentally determined ratio~$N(K)/N_\mathrm{Sp}(K)$ (Fig.~\ref{fig:envelopes}B) as well as the measured contrast~$C(8\hbar k)$: 
\begin{align} \label{eq:atomcontrast}
    C(K) = \frac{N(K)- N_\mathrm{OS}(K)}{N(K)}= 1- \frac{1-C(8\hbar k )}{N(K)/N_\mathrm{Sp}(K)}~.
\end{align}
The green diamonds in Fig.~\ref{fig:contrast} depict the results of Eq.~\eqref{eq:atomcontrast}.

\section*{Section S4. Contrast reduction due to distortions of the twin-lattice beam}

\renewcommand{\thefigure}{S\arabic{figure}}
\renewcommand{\figurename}{Fig.}
\begin{figure}[h]
\begin{center}
\includegraphics[width=.9\linewidth]{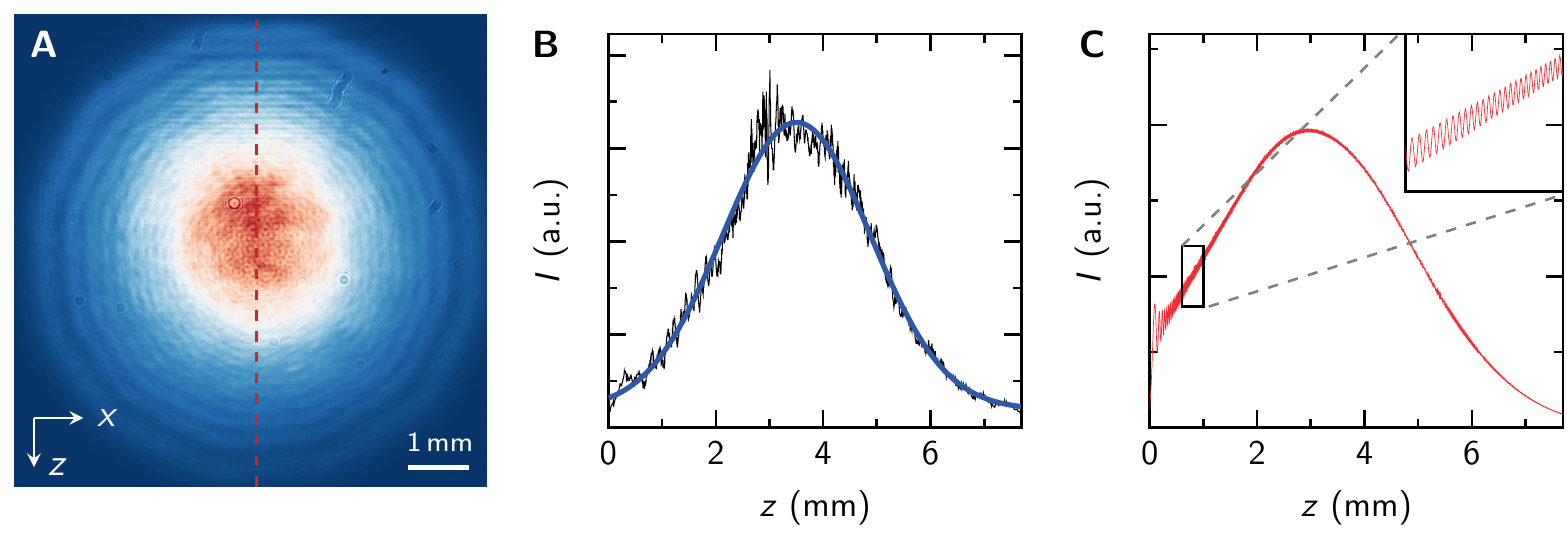}
\caption{\textbf{Measured and simulated twin-lattice beam profile.} (\textbf{A}) Twin-lattice beam imaged with a beam profiling camera after passing through the vacuum chamber. Interference fringes arise due to clipping at apertures and the atom chip. (\textbf{B}) Vertical intensity profile of the beam (black line) along the dashed red line indicated in (\textbf{A}) as well as Gaussian fit to the data (blue line). (\textbf{C}) Simulated intensity profile of a Gaussian beam diffracted at the edge of a metallic half-plane~\cite{Anokhov2004}.} 
\label{fig:beam}
\end{center}
\end{figure}
We study the reduction of the interference contrast due to spatial intensity fluctuations of the twin-lattice laser beam (Fig.~\ref{fig:beam}) deforming the atomic trajectories and causing imperfect mode overlap either in momentum or in position space at $t=2T$. 
 
Experimental observations suggest, that a non-ideal spatial overlap in the direction of the twin lattice can be excluded as it would introduce a timing asymmetry noticeable in the contrast envelopes in Fig.~\ref{fig:envelopes}A.

We investigate the impact of unequal momenta in both interferometer arms, e.g.~$\Delta p_y $ in $y$-direction, which leads to a spatially dependent phase difference~$\delta \varphi_y = \Delta p_y \cdot y / \hbar$, and local interference fringes with spacing~$2\pi\hbar / \Delta p_y$ in the output ports upon spatial imaging. For fringe spacings significantly smaller than the size of the atomic clouds this unwanted phase contribution reduces the measured contrast.

In our model, a path-dependent dipole force acts on the atoms via the gradient of the distorted lattice beam potential
\begin{align}\label{eq:beampotential}
    U(x,y,z) =  V_\mathrm{0}(K) I(x,y,z)/I_0,
\end{align} where $I(x,y,z)/I_0$ is the normalized intensity distribution of the Gaussian lattice beam that has been diffracted at the edge of the atom chip~\cite{Anokhov2004} (Fig.~\ref{fig:beam}C). 
Integration of this dipole force for the interferometer duration $2T$ along each arm reveals a differential momentum between the two arms~$\Delta p_j = p_{j,\mathrm{arm1}} - p_{j,\mathrm{arm2}}$ with~$j=x,y,z$
\begin{align} \label{eq:momentum_trajectory}
\begin{split}
p_{j,\mathrm{arm}}(K) &= - \int_{0}^{2T} \frac{\partial}{\partial j} U(x_\mathrm{arm}(t,K),y_\mathrm{arm}(t,K),z_\mathrm{arm}(t,K))\; \mathrm{d}t\\
&=- \frac{V_0(K)}{I_0} \int_{0}^{2T} \frac{\partial}{\partial j} I(x_\mathrm{arm}(t,K),y_\mathrm{arm}(t,K),z_\mathrm{arm}(t,K))\; \mathrm{d}t~.
\end{split}
\end{align}
$p_{j,\mathrm{arm}}$ depends on the momentum separation $K$ via the wave packet trajectories (Fig.~2) and the lattice depth~$V_\mathrm{0}$ (Fig.~\ref{fig:envelopes}C). 
We evaluate Eq.~(\ref{eq:momentum_trajectory}) for the intensity distribution of the distorted Gaussian beam making the assumption that the atomic motion during the twin-lattice interferometer is given by simplified linearly accelerated trajectories. 

The interferometric contrast $C_\mathrm{LD}$ for a particular value of~$\Delta p_j(K)$ is calculated with the following integral:
\begin{align} \label{eq:LD_contrast}
C_\mathrm{LD} (K) = \left| \iiint \left| \Psi_{ \left(0\hbar k,\pm 2\hbar k\right)}(x,y,z,t=2T)\right|^2 \; e^{-\frac{i}{\hbar}(\Delta p_x \cdot x+\Delta p_y \cdot y+\Delta p_z \cdot z)} \mathrm{d}x \mathrm{d}y \mathrm{d}z \right| .
\end{align}
Since the twin-lattice laser beam is well collimated and there is very little atomic motion in the $x$-direction compared to both the $y$- and $z$-directions, we assume the beam profile to be symmetric in $x$-direction simplifying our calculations by setting $\Delta p_x = 0$.

We model the density $\left|\Psi_{ \left(0\hbar k,\pm 2\hbar k\right)}(x,y,z,t=2T)\right|^2$ of the two interfering wave packets at the interferometer output ports with momenta~$ 0\hbar k$ and~$\pm 2\hbar k$ by using Thomas-Fermi density distributions. Their sizes have been inferred from time-of-flight measurements. In order to account for a dependence of~$\Delta p_j(K)$ on the spatial extent of the atomic wave packet we calculate a sample of single-particle trajectories with different initial positions given by density distribution at the beginning of the interferometer sequence~$\left|\Psi_{\mathrm{I}}(x,y,z,t=0)\right|^2$  and average the contrast values in Eq.~\eqref{eq:LD_contrast} over those realizations providing an average light-diffraction contrast $C_\mathrm{LD,avg} (K)$.

To match our simplified model with the experiment, we use a single fitting parameter, namely a factor multiplying the normalized intensity ratio in Eqs.~\eqref{eq:beampotential} and \eqref{eq:momentum_trajectory}. Our simplified light distortion model features intensity perturbations $|I(x,y,z)-I_\mathrm{Gauss}|/I_0$ of about 1\% caused by the diffraction on the metallic edge (Fig.~\ref{fig:beam}C). 
The contrast values (red triangles) depicted in Fig.~\ref{fig:contrast} have been obtained enhancing the perturbations to 9\% providing good agreement with the experiment. 
One has to note, that assuming a beam clipped on a single chip edge oversimplifies the experimental situation, where diffraction occurs at different apertures causing larger total distortions. 
A 3D simulation of the totality of these diffractions requires a more detailed study and goes beyond the scope of this work. 
Direct measurements of the beam profile outside of the vacuum chamber as depicted in Fig.~\ref{fig:beam}A, however, show intensity variations at the level of 10\% and, thus, support our assumptions.
In turn, increasing the intensity of an ideal Gaussian beam does not lead to a significant loss of signal in our model. To maintain a contrast of $C_\mathrm{LD,avg} = 90$\% up to $K=408{\hbar}k$ the intensity fluctuations should not exceed 0.5\%.
The confidence intervals in Fig.~\ref{fig:contrast} reflect the 10\% uncertainty in the measured lattice depth~$V_\mathrm{0}(K)$.
The orange pentagons represent the product of the results of Eqs.~\eqref{eq:atomcontrast} and \eqref{eq:LD_contrast},~$C_{\mathrm{total}}(K) =C_\mathrm{LD,avg}(K) \cdot C(K)$.

\begin{figure}[p]
\begin{center}
\includegraphics[width=0.9\linewidth]{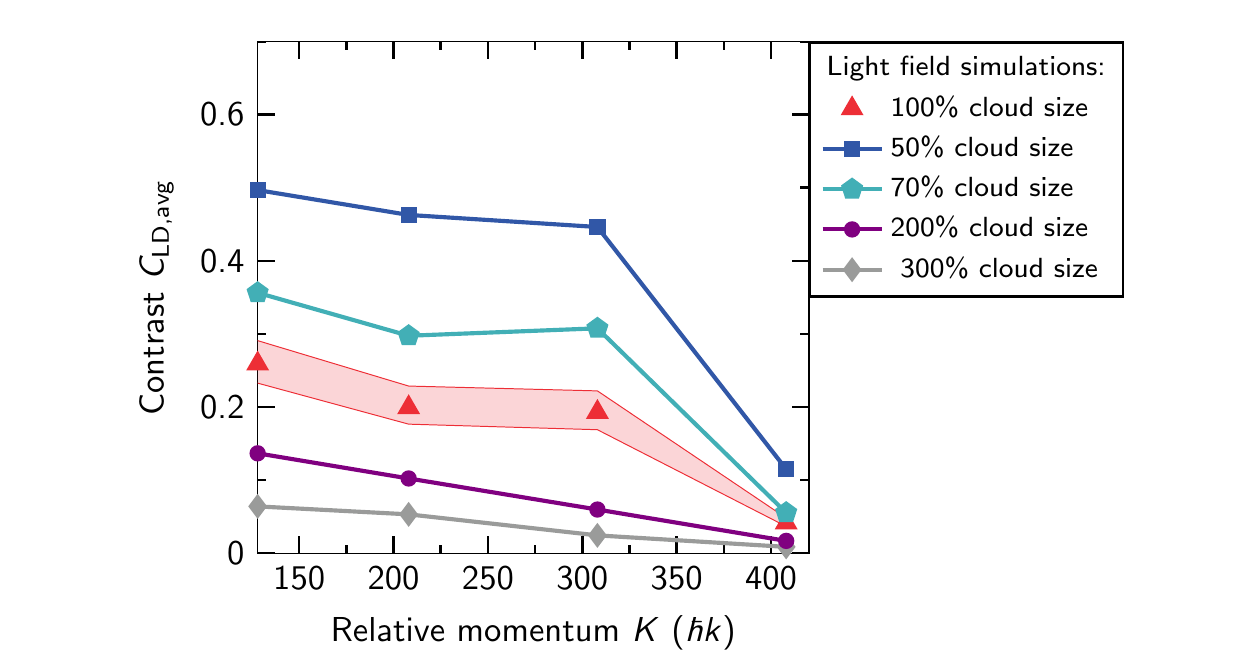}
\caption{\textbf{Simulated contrast $\mathbf{C_\mathrm{LD,avg}(K)}$ in dependence of cloud size.} We show the calculated contrast reduction due to the interaction of the BEC with the distorted twin-lattice beam according to Eq.~\eqref{eq:LD_contrast} (red triangles in Fig.~\ref{fig:contrast}) for large relative momenta $K>100\hbar k$. 
We compare the contrast $C_\mathrm{LD,avg}(K)$ of our delta-kicked collimated BEC (100\% cloud size, red triangles) to values calculated for different cloud sizes but otherwise identical input parameters. 
A magnification of the cloud's spatial spread to 200\% or 300\% (violet circles, gray diamonds) increases the phase variation across the cloud and, therefore, leads to a significant contrast reduction.
Clouds with smaller radii of $70\%$ (turquoise pentagons) and $50\%$ (blue squares) in turn provide improved contrast highlighting the benefit of a well-collimated atom source in the presence of light field distortions.
}
\label{fig:CloudSizeContrast}
\end{center}
\end{figure}

\renewcommand{\thefigure}{S\arabic{figure}}
\begin{figure}[htbp]
\begin{center}
\includegraphics[width=0.9\linewidth]{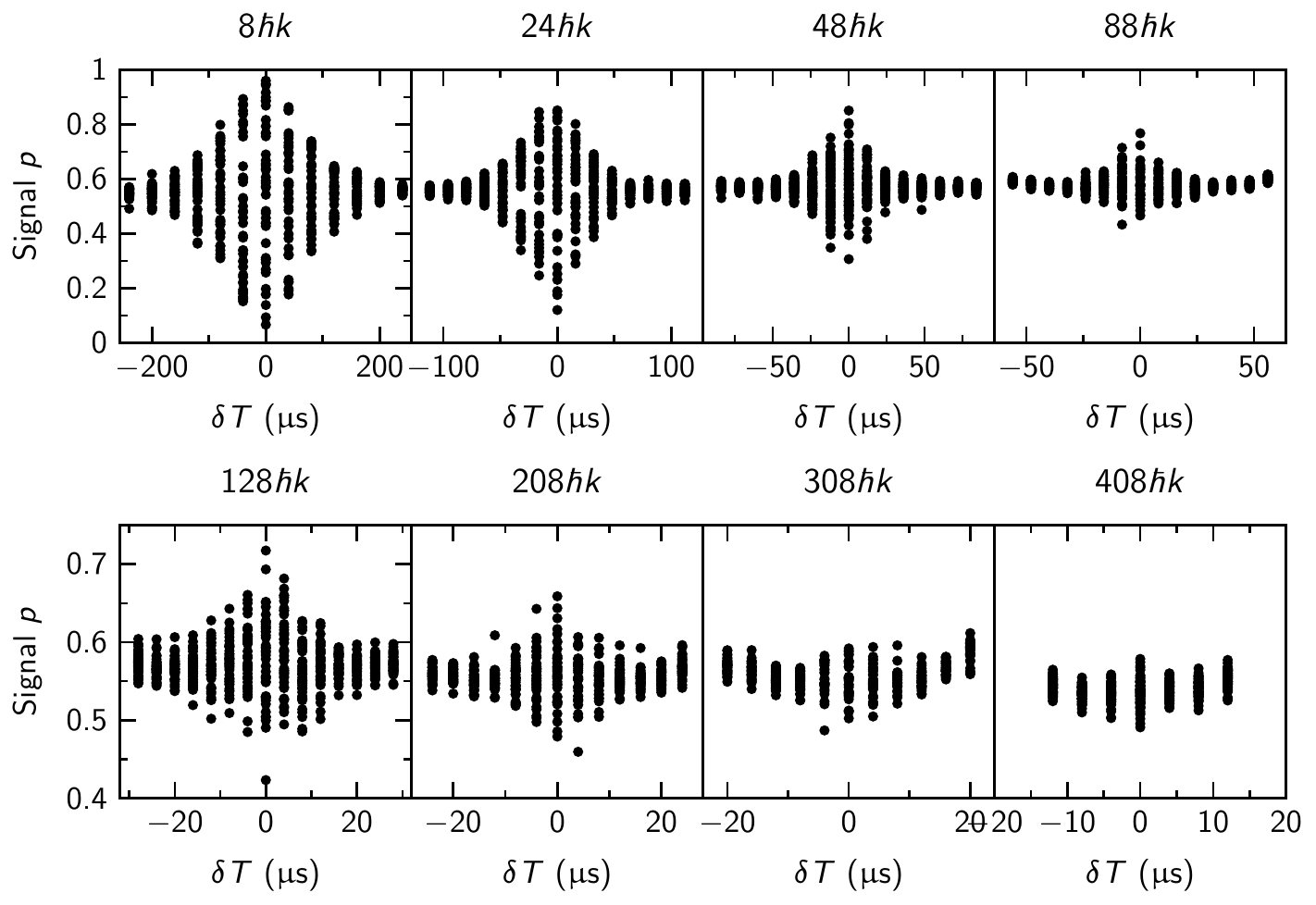}
\caption{\textbf{Measured interferometer signal $p$ as a function of asymmetry $\delta T$.} 
We plot the signal $p=(N_{+2{\hbar}k}+N_{-2{\hbar}k})/(N_{+2{\hbar}k}+N_{-2{\hbar}k}+N_{0{\hbar}k})$ depending on $\delta T$ for different relative momenta $K=(8,24,48,88,128,208,308,408)~\hbar k$ in the twin-lattice interferometer. For each $\delta T$ we take 40 data points to calculate the standard deviation $\sigma_p(\delta T)$ of $p$ (Fig.~\ref{fig:envelopes}A).}
\label{fig:suppl:AI_signal}
\end{center}
\end{figure}

\renewcommand{\thetable}{S\arabic{table}}


\begin{center}
\begin{table}[h]
\begin{tabular}{ |l|l|c|c|c|c| } 
 \hline
  & Type & $A$ (mm$^2$) & $\tau$ (ms) & $L$ (mm) & Calculation of $A$ and $L$ \\\hline
 Twin lattice & & 7.6 & 12.1 & 2.43  & $L=\frac{1}{2}g((\tau + t_0)^2 - t_0^2)$ \\ 
 Savoie~\cite{Savoie18SciAdv}  & Butterfly & 1120 & 801 & 787 & $A=\frac{1}{2}v_\mathrm{r,Cs}g(\tau/2)^3$, $L=\frac{1}{2}g (\tau/2)^2$ \\ 
 Stockton~\cite{Stockton11PRL}  & Butterfly & 19 & 206 & 52  & $A=\frac{1}{2}v_\mathrm{r,Cs}g(\tau/2)^3$, $L=\frac{1}{2}g (\tau/2)^2$ \\ 
 Berg~\cite{Berg15PRL} & Mach-Zehnder & 41 & 50 & 140  & $A=\tau^2 vv_\mathrm{r,Rb}$, $L=\tau v$ \\ 
 Canuel~\cite{Canuel06PRL} & Mach-Zehnder & 2.1 & 60 & 19.8  & $A=2(\tau/2)^2 vv_\mathrm{r,Cs}$, $L=\tau v$ \\
 Moan.~\cite{Moan20PRL} & Ring-shaped guide & 0.5 & 107.7 & 0.4  & $A=4\pi R^2$, $L=2R$ \\
 Pandey~\cite{Pandey19Nat}$^\ast$ & Ring-shaped guide & 25.3 & 2000 & 0.886  & $A=41\cdot \pi R^2$, $L=2R$ \\
 Wu~\cite{Wu07PRL} & Moving guide & 0.18 & 50 & 1  & $A = L/\pi 2v_\mathrm{r} \tau $ \\
 \hline
\end{tabular}
\\ $^\ast$ No realization of an interferometer
\caption{\textbf{Comparison of areas enclosed by different Sagnac interferometers}.
Parameters used to calculate the area $A$ and compactness factor $(\tau L)^{-1}$ in Fig.~\ref{fig:gyroscopes}. Whenever possible, a formula is given for the calculation of the area $A$ and the baseline $L$.
$t_0$ is the free-fall time before the interferometer,  $v_\mathrm{r,Rb/Cs}=\hbar k/m_\mathrm{Rb/Cs}$ the recoil velocity and $v$ the forward velocity perpendicular to both gravity $g$ and the wave vector $k$.\\ \\
1.  D. Savoie \textit{et al}. Interleaved atom interferometry for high-sensitivity inertial measurements.  \textit{Sci. Adv.} \textbf{4} (2018).\\
24.  J. Stockton \textit{et al}. Absolute geodetic rotation measurement using atom interferometry. \textit{Phys. Rev. Lett.} \textbf{107}, 133001 (2011).\\
12.  P. Berg \textit{et al}. Composite-light-pulse technique for high-precision atom interferometry. \textit{Phys. Rev. Lett.} \textbf{114}, 063002 (2015).\\
23.  B. Canuel \textit{et al}. Six-axis inertial sensor using cold-atom interferometry. \textit{Phys. Rev. Lett.} \textbf{97}, 010402 (2006).\\
26.  E. R. Moan \textit{et al}. Quantum rotation sensing with dual sagnac interferometers in an atom-optical waveguide. \textit{Phys. Rev. Lett.} \textbf{124}, 120403 (2020).\\
30.  S. Pandey \textit{et al}. Hypersonic Bose-Einstein condensates in accelerator ring. \textit{Nature} \textbf{570}, 205-209 (2019).\\
25.  S. Wu \textit{et al}. Demonstration of an area-enclosing guided-atom interferometer for rotation
sensing. \textit{Phys. Rev. Lett.} \textbf{99}, 173201 (2007).
}
\end{table}
\end{center}

\end{document}